%% file: DRL_LIS_SPAWC_AA_V59A.tex
\documentclass[conference]{IEEEtran}
\input{input.tex}
\usepackage{epstopdf}
\usepackage{enumerate}
\usepackage{algorithm}
\usepackage{amsmath}
\usepackage{amssymb,amsfonts}
\usepackage[noend]{algpseudocode}
\usepackage{float}
\usepackage{color}
\usepackage{makeidx}
\usepackage{bbm}
\usepackage{graphicx}
\usepackage{cleveref}
\usepackage{url}
\usepackage{steinmetz}
\usepackage{varwidth}
\usepackage{textcomp}
\usepackage{xcolor}
\usepackage{balance}

\newcommand{\sref}[1]{{Section}~\ref{#1}}

\DeclareMathOperator*{\argmax}{arg\,max}

\newcommand{\tc}[1]{\textcolor{cyan}{#1}}

\def \bEpsi{\boldsymbol{\Psi}}

\def \rm {\mathrm}
\def \bpsi{\boldsymbol{\psi}}

\algnewcommand{\Initialize}[1]{%
	\State \textbf{Initialization:} \parbox[t]{.8\linewidth}{\raggedright #1}}

\begin{document}
\title{Deep Reinforcement Learning for  Intelligent  Reflecting Surfaces: Towards Standalone Operation}
\author{\IEEEauthorblockN{Abdelrahman Taha$^\star$, Yu Zhang$^\star$, Faris B. Mismar$^\dag$, and Ahmed Alkhateeb$^\star$}
	\IEEEauthorblockA{$^\star$\textit{Arizona State University}, Tempe, AZ, USA, 	Emails: $\{$a.taha, y.zhang, aalkhateeb$\}$@asu.edu}
	$^\dag$\textit{The University of Texas at Austin}, Austin, TX, USA, Email: {faris.mismar@utexas.edu}
}
\maketitle

\begin{abstract}
The promising coverage and spectral efficiency gains of intelligent reflecting surfaces (IRSs) are attracting increasing interest. In order to realize these surfaces in practice, however, several challenges need to be addressed. One of these main challenges is how to configure the reflecting coefficients on these passive surfaces without requiring massive channel estimation or beam training overhead. Earlier work suggested leveraging supervised learning  tools to design the IRS reflection matrices. While this approach has the potential of reducing the beam training overhead,  it requires collecting large datasets for training the neural network models. In this paper, we propose a novel deep reinforcement learning framework for predicting the IRS reflection matrices with minimal training overhead.  Simulation results show that the proposed online learning framework can converge to the optimal rate that assumes perfect channel knowledge. This represents an important step towards realizing a \textit{standalone IRS operation}, where the surface configures itself without any control from the infrastructure.  
\end{abstract}

\begin{IEEEkeywords}
	reconfigurable intelligent surface, large intelligent surface, intelligent reflecting surface, smart reflect-array, beamforming, deep reinforcement learning
\end{IEEEkeywords}

\section{Introduction} \label{sec:Intro}

The increasing demand on data rates from the massive number of devices motivates the need to develop novel system architectures that are both energy and spectrally efficient.  For the past years, state of the art research has focused on leveraging large-scale MIMO systems, such as massive and millimeter wave (mmWave) MIMO at the base stations (BSs) and mobile users. To further improve the coverage and the energy efficiency of these systems, intelligent reflecting surfaces (IRSs) have been recently proposed and attracted massive interest\cite{Huang2019,Taha2019,Basar2019b,Liang2019,Yuan2020}. 
%
%
IRSs consist of a huge number of passive reflecting elements whose function is to reflect the incident signal \textit{intelligently} into the desired directions, by means of software-controllable phase shifts. Since the IRS reflection beamforming design requires the perfect/imperfect channel knowledge, the channel estimation is a crucial aspect for the IRS interaction design problem. The massive number of passive IRS elements, however, impose a main challenge on acquiring the channel estimates; traditional channel estimation solutions will lead to either huge training overhead or prohibitive hardware complexity for the IRS architectures \cite{Taha2019}. Given an end goal of achieving harmonic co-existence between all the heterogeneous wireless systems, setting an objective of developing \textit{fully-standalone} IRS architectures seems as the next step forward for reaching that end goal. 

Prior work focused on proposing solutions for both the channel estimation and the reflection beamforming design problems \cite{Taha2019,Huang2019a,Elbir2020,Jensen2019}. 
The authors in \cite{Taha2019} proposed the first solution to the channel/beam training overhead challenge leveraging tools from both compressivse sensing and supervised deep learning. The promising gains of these solutions motivated more research in these directions. For example, in \cite{Elbir2020}, a supervised deep learning framework is used for channel estimation by mapping the received pilots to the direct and the cascaded channels. In \cite{Jensen2019}, an IRS channel estimation scheme based on a minimum variance unbiased estimator is proposed. The solutions in \cite{Taha2019,Elbir2020,Jensen2019}, however, either considered supervised deep learning which  requires large dataset collection phase before training, or  assumed that the IRS is assisted/controlled by another base station/access point, not operating on its own.

This work presents a \textit{novel} application of deep reinforcement learning in predicting the reflection coefficients of the IRS surfaces without requiring any prior training overhead.  
The main contributions of this paper can be summarized as follows.
\begin{itemize}
	\item \emph{A novel deep reinforcement learning (DRL) based solution} is proposed for the IRS interaction design, where the IRS learns how to reflect the incident signals in the best possible way by adjusting its reflection matrix. This solution eliminate the need for collecting large training dataset, hence requires almost no training overhead. 
	
	\item The proposed framework is directed more towards \emph{standalone IRS operation}, where the IRS architecture is not controlled/assisted by any base station, but rather operating on its own while interacting with the environment, and without any initial training phase requirement.
\end{itemize}
 Simulation results based on accurate 3D ray-tracing datasets show that the achievable rates of the proposed DRL based solution can converge close to the upper bound with an added value of almost no training overhead, as opposed to supervised learning based solutions.

\begin{figure*}[t] \centerline{\includegraphics[width=1.8\columnwidth]{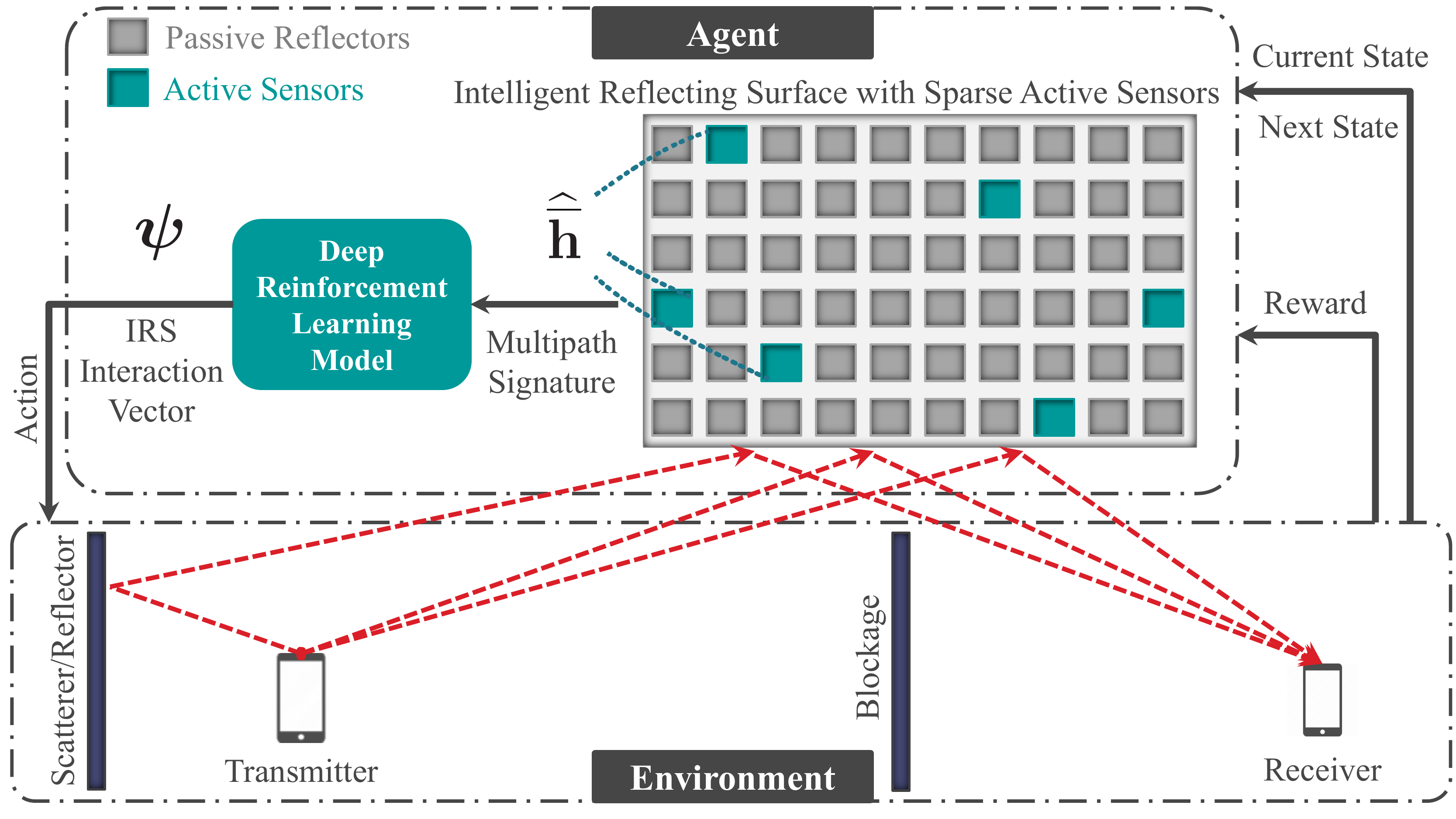}} 
	\caption{The transmitter-receiver communication is assisted by a intelligent reflecting surface (IRS). The IRS is interacting with the incident signal through an interaction vector $\boldsymbol{\psi}$. Active channel sensors are randomly distributed over the IRS. These active elements have two modes of operation (i) a channel sensing mode where it is connected to the baseband to estimate the channels and (ii) a reflection mode where it just reflects the incident signal by applying a phase shift. The rest of the IRS elements are passive reflectors. The environment is represented by the various scatterers, user locations, etc ... The IRS acts as a reinforcement learning agent by acquiring a state and a reward from the environment and exerting an action back on the environment.}
	\label{fig:Sys_Model}
\end{figure*}

\textbf{Notation:} $\bA$ is a matrix, $\ba$ is a vector, $a$ is a scalar, and $\boldsymbol{\mathcal{A}}$ is a set of vectors. $\mathrm{diag}(\ba)$ is a diagonal matrix with entries of $\ba$ on its diagonal. $|\bA|$ is the determinant of $\bA$, $\bA^T$ is its transpose, $[\bA]_{r,:}$ is the $r^\rm{th}$ row of $\bA$,  and $\boldsymbol{\mathrm{vec}}(\mathbf{A})$ is a vector whose elements are the stacked columns of $\mathbf{A}$.  $\bI$ is the identity matrix. $\bA \odot \bB$ is the Hadamard product of $\bA$ and $\bB$. $\cN(\bm,\bR)$ is a complex Gaussian random vector with mean $\bm$ and covariance $\bR$. $\bbE\left[\cdot\right]$ is for expectation. 

\section{System and Channel Models} \label{sec:SysCh Model}


\subsection{System Model} \label{sec:Sys_Model}
Consider an OFDM-based system of $K$ subcarriers where a single-antenna transmitter is communicating with a single-antenna receiver due to the assistance of an $M$-elements intelligent reflecting surface (IRS), as in \figref{fig:Sys_Model}. Let $\bh_{\rm{T},k}, \bh_{\rm{R},k} \in \mathbb{C}^{M \times 1}$ denote the channels from the transmitter/receiver to the IRS at the $k^\rm{th}$ subcarrier. $s_k$ is the transmit signal, where $\bbE[\left|s_k\right|^2]=\frac{P_\rm{T}}{K}$. $P_\rm{T}$ is the total transmit power. $\bEpsi$ denotes the IRS interaction diagonal matrix. $n_k \sim \mathcal{N}_\mathbb{C}(0,\sigma_n^2)$ is the receive noise. The receive signal at the receiver can be expressed as
\begin{align}
y_k&=  \bh_{\rm{R},k}^T \bEpsi \ \bh_{\rm{T},k} s_k + n_k, \\
&\stackrel{(a)}{=}  \left( \bh_{\rm{R},k} \odot  \bh_{\rm{T},k}  \right)^T \boldsymbol{\psi} \ s_k + n_k,
\end{align}
where $\boldsymbol{\psi}$ is the IRS interaction vector, such that $\bEpsi =\mathrm{diag}\left({\boldsymbol{\psi}}\right)$. Assume an IRS architecture of RF phase shifters, every interaction factor can be represented as $\left[\boldsymbol{\psi}\right]_m=e^{j \phi_m}$, hence the choice of an interaction vector is constrained to a predefined codebook $\boldsymbol{\mathcal{P}}$. Adopting the IRS architecture proposed in \cite{Taha2019} and illustrated in \figref{fig:Sys_Model}, \textit{active} elements are randomly distributed over the IRS. The \textit{sampled} channel vector from the transmitter/receiver to the IRS active elements, $\overline{\bh}_{\rm{T},k},\overline{\bh}_{\rm{R},k} \in \mathbb{C}^{\overline{M} \times 1}$, can be expressed as $\overline{\bh}_{\rm{T},k} = \bG_\rm{IRS} \bh_{\rm{T},k}$ and $ \overline{\bh}_{\rm{R},k} = \bG_\rm{IRS} \bh_{\rm{R},k}$, where $\bG_\rm{IRS}$ is an $\overline{M} \times M$ selection matrix that selects the entries corresponding to the active IRS elements. Finally, the overall IRS \textit{sampled} channel vector can be expressed as $\overline{\mathbf{h}}_{k} = \overline{\mathbf{h}}_{\rm{T},k} \odot \overline{\mathbf{h}}_{\rm{R},k}$. 

\subsection{Channel Model} \label{subsec:Ch_Model}
A wideband gemoetric channel model is adpoted \cite{Alkhateeb2018}. Consider a transmitter-IRS channel, $\bh_{\rm{T},k}$, (and similarly for the IRS-receiver channel) consisting of $L$ clusters. Each cluster contributes with one ray from the transmitter to the IRS. The ray parameters are: azimuth/elevation angles of arrival, $\theta_{\ell},\phi_{\ell} \in [0, 2 \pi)$; complex coefficient $\alpha_{\ell} \in \bbC$; time delay $\tau_{\ell} \in \bbR$. The transmitter-IRS path loss is denoted by $\rho_{\rm{T}}$. The pulse shaping function, with $T_{S}$-spaced signaling, is defined as $p\left(\tau\right)$ at $\tau$ seconds. 
The frequency domain channel vector, $\bh_{\rm{T},k}$, can then be defined as 

\begin{equation}
\bh_{\rm{T}, k} = \sqrt{\frac{M}{\rho_\rm{T}}} \sum_{d=0}^{D-1} \sum_{ \ell =1}^L \alpha_{\ell} \ \ba\left(\theta_{\ell}, \phi_{\ell} \right)  p(d T_S - \tau_{\ell}) \ e^{-j\frac{2\pi k}{K}d},
\label{eq:Ch_k}
\end{equation} 

where $\ba(\theta_{\ell},\phi_{\ell}) \in \mathbb{C}^{M \times 1}$ is the IRS array response vector. Assume a block-fading channel model, where $\bh_{\rm{T},k}$ and $\bh_{\rm{R},k}$ are assumed to stay constant over the channel coherence time. 

\section{Problem Formulation} \label{sec:Formulation}

Given the objective of maximizing the achievable rate at the receiver, our problem is then to find the optimal interaction vector, $\bpsi^\star$, that solves

\begin{equation} \label{eq:optV}
\begin{split}
& \bpsi^\star=  \argmax_{\bpsi \in \boldsymbol{\mathcal{P}}} \sum_{k=1}^K  \log_2\left(1+\mathsf{SNR} \left| \left(\bh_{\rm{T},k} \odot \bh_{\rm{R},k}\right)^T \boldsymbol{\psi} \right|^2 \right), 
\end{split}
\end{equation}

to achieve the optimal rate $R^\star$ defined as
\begin{equation} \label{eq:optR}
\begin{split}
& R^\star= \frac{1}{K} \sum_{k=1}^K  \log_2\left(1+\mathsf{SNR} \left| \left(\bh_{\rm{T},k} \odot \bh_{\rm{R},k}\right)^T \boldsymbol{\psi}^\star \right|^2 \right). 
\end{split}
\end{equation}

Unfortunately, there is no closed form solution for the optimization problem in \eqref{eq:optV} due to the quantized codebook constraint and the use of one interaction vector $\bpsi$ fixed over all subcarriers. Accordingly, finding the optimal interaction vector for the IRS, $\bpsi^\star$, requires an exhaustive search over the codebook $\boldsymbol{\mathcal{P}}$.
This search, however, leads either to prohibitive training overheard, hardward complexity, or power consumption, as detailed in \cite{Taha2019}. Our objective is then to find an efficient solution for the IRS systems that approaches the optimal rate in \eqref{eq:optR} with \textbf{almost no training overhead} and with an \textbf{energy-efficient hardware}. In the next section, we propose a \textit{novel} application of deep reinforcement learning in the interaction design problem of intelligent reflecting surfaces. This solution actually eliminates the need for collecting large training datasets as opposed to the supervised learning solution proposed in \cite{Taha2019}. The supervised learning solution, however, approaches the optimal rate with fewer iterations.

\section{Deep Reinforcement Learning Based IRS Interaction Design}  \label{sec:DL_Solution}

\subsection{Key Idea} \label{subsec:keyIdea}

From \eqref{eq:optV}, the optimal interaction vector is a function of the channels between the two communication ends and the IRS. To avoid the prohibitive overhead of estimating the full IRS channels, the optimal interaction vector choice can be mapped to the surrounding \textit{environment}, which the full IRS channels inherently describe. Modeling the various elements of the environment, mathematically, is notoriously complicated. In contrast, leveraging an awareness of the environment using a multipath signature \cite{Alkhateeb2018} can be sufficient. In such case, deep reinforcement learning models can be adopted to learn the mapping function from multipath signatures to the optimal interaction vectors as illustrated in \figref{fig:Sys_Model}. The IRS active elements play a crucial role in capturing one form of mutlipath signatures: the \textit{sampled} channels, $\overline{\bh}_{\rm{T},k}, \overline{\bh}_{\rm{R},k}$.
Fortunately, estimating the sampled channel vectors can be accomplished with a few pilot signals; i.e., negligible training overhead. This solution also involves energy-efficient low-complexity hardware architectures (few sparse active IRS elements) \cite{Taha2019}.

\subsection{Proposed Solution} \label{subsec:Sys_Oper}
The proposed deep reinforcement learning (DRL) based IRS interaction design approach operates in two parts: (I) the agent interaction and (II) the agent learning, as in Algorithm \ref{alg1}. The IRS interchanges between these two parts continuously.

\textbf{PART I: Agent Interaction}

The IRS interaction with the \textit{environment} can be outlined as follows: the IRS observes the current \textit{state}, $s$, of the environment and takes an \textit{action}, $a$, predicated upon the observed state. The IRS then receives a \textit{reward}, $r$, for the action taken and a new \textit{state} observation, $s'$, from the environment. Once the experience is acquired, $\left\langle s,a,r,s' \right\rangle$, the IRS trains the DRL model using current and past experiences, in the second part.

Let the term ``experience" indicates the information captured in one learning episode, and define the concatenated \textit{sampled} channel vector as

\begin{align}
\overline{\mathbf{h}} = \textbf{vec} \left(  \left[ \overline{\mathbf{h}}_{1}, \overline{\mathbf{h}}_{2}, \dotsc, \overline{\mathbf{h}}_{K} \right] \right).
\label{eq:noisy_ch}
\end{align}

 Assume that the one learning episode occurs every coherence block and let $T$ be the maximum number of episodes, $\overline{\mathbf{h}}(t)$ denotes the concatenated \textit{sampled} channel vector at the $t^\rm{th}$ episode, where $t=1, ..., T$. Part I steps are summarized as follows.
 
 \textbf{\emph{1. Sampled channel estimation (lines 3,13):}} The transmitter and receiver transmits two orthogonal uplink pilots. The IRS active elements will receive these pilots and estimate the \textit{sampled} channel vectors to construct the multipath signature. 
 
 \begin{align}
 & \widehat{\overline{\bh}}_{\rm{T}, k}(t)  =  \overline{\bh}_{\rm{T},k}(t) + \bv_k, \widehat{\overline{\bh}}_{\rm{R}, k}(t) =  \overline{\bh}_{\rm{R},k}(t) + \bw_k, \\
 & \widehat{\overline{\mathbf{h}}}_{k}(t) = \widehat{\overline{\mathbf{h}}}_{\rm{T},k}(t) \odot \widehat{\overline{\mathbf{h}}}_{\rm{R},k}(t), \\
 & \widehat{\overline{\mathbf{h}}}(t) = \textbf{vec} \left(  \left[ \widehat{\overline{\mathbf{h}}}_{1}(t), \widehat{\overline{\mathbf{h}}}_{2}(t), \dotsc, \widehat{\overline{\mathbf{h}}}_{K}(t) \right] \right).
 \label{eq:noisy_ch}
 \end{align}
 where $\bv_k,\bw_k \sim \mathcal{N}_\mathbb{C} \left(\boldsymbol{0}, \sigma_n^2 \bI \right)$ are the receive noise vectors.
 
 \textbf{\emph{2. Data transmission (lines 5-10):}}
 The multipath signature is used to predict the interaction vector. To account for exploration (i.e., randomly sampling from the action space) besides exploitation (i.e., using prior learning experience), the factor $\epsilon$ is introduced such that an interaction vector can be randomly chosen out of the codebook $\boldsymbol{\cP}$ with $\epsilon$ probability. Otherwise, the interaction vector is predicted from the current network. After that, the interaction vector chosen, reflects the transmitted data from the transmitter.
 
 \textbf{\emph{3. Feedback reception (lines 11,12):}}
 The IRS receives a feedback from the receiver indicating the achievable rate, $R(t)$, attained by using the interaction vector, which is defined as
 \begin{equation} \label{eq:23}
 \begin{split} 
 & R(t) = \\
 & \hspace{-10pt} \frac{1}{K}  \sum_{k=1}^K  \log_2\left(1+\mathsf{SNR} \left| \left(\bh_{\rm{T},k}(t) \odot \bh_{\rm{R},k}(t)\right)^T \boldsymbol{\psi}_a \right|^2 \right).
 \end{split}
 \end{equation} 
 \normalsize
 After that, the rate is quantized based on a threshold level, such that $R_\rm{Q}(t) = 1$ if $R(t) > R^\rm{TH}$; otherwise, $R_\rm{Q}(t) = -1$. Reward clipping is substantial for learning convergence \cite{Mnih2015}.

\begin{algorithm}[h]
	\caption{Deep Reinforcement Learning Based IRS Interaction Design}
	\label{alg1}
	\begin{algorithmic}[1]
		\small 
		\Require{Reflection beamforming codebook $\boldsymbol{\cP}$.} 
		\Ensure{Trained network $Q\left( s,a|\boldsymbol{\theta}\right)$.}
		\Initialize{ Network $Q\left( s,a|\boldsymbol{\theta}\right)$, replay buffer $\cD$.}
		
		\Repeat
		
		\State \begin{varwidth}[t]{\linewidth} 
			IRS receives two pilots to estimate $\widehat{\overline{\bh}}(1)$. \Comment{\tc{Current state}}
		\end{varwidth}
		
		\For{episode $t=1$ \textbf{to} $T$} \Comment{\tc{For every episode}}
		
		\Statex \textbf{\hspace{\algorithmicindent}\hspace{\algorithmicindent}PART I: Agent Interaction} \nonumber
		
		\State \begin{varwidth}[t]{\linewidth}
			Sample $\xi \sim \mathrm{Uniform}\left( 0,1\right) $ 
		\end{varwidth}
		
		\If{$\xi \leq \epsilon$} \Comment{\tc{Select action}} 
		\State \begin{varwidth}[t]{\linewidth} \footnotesize Select interaction vector, $\boldsymbol{\psi}(t) \in \boldsymbol{\cP}$ at random. \small  \end{varwidth} 
		\Else 
		\State \begin{varwidth}[t]{\linewidth} \footnotesize Select interaction vector, $\boldsymbol{\psi}(t) =\argmax_{a'} Q\left( s,a'|\boldsymbol{\theta}\right)$. \small \end{varwidth} 
		\EndIf
		
		\State \begin{varwidth}[t]{\linewidth}
			IRS reflects using $\boldsymbol{\psi}(t)$ beam. \Comment{\tc{Carry out action}}	 	
		\end{varwidth}
		
		\State \begin{varwidth}[t]{\linewidth}
			IRS receives the feedback $R(t)$.	\Comment{\tc{Observe reward}}	 
		\end{varwidth}
		
		\State \begin{varwidth}[t]{\linewidth}
			IRS quantizes the reward, $R_\rm{Q}(t) \in \left\lbrace\pm 1 \right\rbrace $. 
		\end{varwidth}
		
		\State \begin{varwidth}[t]{\linewidth} 
			IRS receives two pilots to estimate $\widehat{\overline{\bh}}(t+1)$ \Comment{\tc{Next state}}
		\end{varwidth}
		
		\Statex \textbf{\hspace{\algorithmicindent}\hspace{\algorithmicindent}PART II: Agent Learning} \nonumber
		\State \begin{varwidth}[t]{\linewidth} 
			$\left\langle s, a, r, s' \right\rangle  \gets \left\langle \widehat{\overline{\bh}}(t), \boldsymbol{\psi}(t), R_\rm{Q}(t), \widehat{\overline{\bh}}(t+1) \right\rangle$.
		\end{varwidth}
		
		\State \begin{varwidth}[t]{\linewidth} 
			Store the experience $\left\langle s, a, r, s' \right\rangle$ in $\cD$.
		\end{varwidth}
		
		\State \begin{varwidth}[t]{\linewidth} 
			Minibatch experiences from $\cD$ for training.
		\end{varwidth}
		
		\State \begin{varwidth}[t]{\linewidth} 
			feedforward $s$ to calculate $\widehat{\bR}(t) \gets Q\left( s,a|\boldsymbol{\theta}\right) \forall a$.
		\end{varwidth}
		
		\State \begin{varwidth}[t]{\linewidth} 
			feedforward $s'$ to calculate $\Gamma \gets \max_{a'} Q\left( s',a'|\boldsymbol{\theta}\right)$ \\and calculate $a^\star \gets \argmax_{a'} Q\left( s',a'|\boldsymbol{\theta}\right)$.
		\end{varwidth}
		
		\State \begin{varwidth}[t]{\linewidth} 
			Construct the target vector, $\overline{\bR}(t)$:
		\end{varwidth}
		
		\State \begin{varwidth}[t]{\linewidth} \footnotesize $\left[ \overline{\bR}(t)\right]_{a^\star} \gets R_\rm{Q}(t) + \gamma \Gamma $, \small  \end{varwidth} 
		
		\State \begin{varwidth}[t]{\linewidth} \footnotesize $\left[ \overline{\bR}(t)\right]_{a' \neq a^\star} \gets \left[ \widehat{\bR}(t)\right]_{a' \neq a^\star}, a' \in \left\lbrace 1,\dotsc,\left| \boldsymbol{\cP} \right|  \right\rbrace$. \small  \end{varwidth} 
		
		\State \begin{varwidth}[t]{\linewidth} 
			Perform SGD on $\mathsf{MSE} \left( \overline{\bR}(t), \widehat{\bR}(t)\right)$ to find $\boldsymbol{\theta}^\star$.
		\end{varwidth}
		
		\State \begin{varwidth}[t]{\linewidth} 
			Update network weights $\boldsymbol{\theta}(t) \gets \boldsymbol{\theta}^\star$.
		\end{varwidth}
		
		\State \begin{varwidth}[t]{\linewidth} 
			Decrease $\epsilon$ gradually.
		\end{varwidth}	 
		
		\State \begin{varwidth}[t]{\linewidth} 
			$s \gets s'$. \Comment{\tc{Assign next state to current state}}
		\end{varwidth}
		\normalsize	
		\EndFor
		\Until{reaching a terminal goal}
	\end{algorithmic}
\end{algorithm}

\textbf{PART II: Agent Learning}

The IRS leverages the acquired experiences to train the DRL model. Part II steps are summarized as follows.

\textbf{\textit{1. Constructing a new experience (lines 14,15):}}
The new experience acquired is now stored in the experience replay buffer $\cD$ for training of the deep Q-network \cite{Mismar2019}.

\textbf{\emph{2. Model training (lines 16-23):}} The deep Q-network is now trained to minimize the prediction loss. To do so, we use the stochastic gradient descent algorithm (SGD). The training operates sequentially using minibatchs from the replay buffer $\cD$. It learns how to map an input state (\textit{sampled} channel vector) to an output action (interaction vector).

\subsection{Machine Learning Design} \label{subsec:DL Model}

\textbf{\textbullet\ Input Representation:} the concatenated \textit{sampled} channel vector, $\widehat{\overline{\bh}}$, is the input to the deep Q-network. The normalization method used is a simple per-dataset scaling \cite{Zhang2019,Li2019}; all samples are normalized by the maximum absolute value over the whole input data. This method preserves distance information encoded in the multipath signatures. Each complex entry of the input data is split into real and imaginary values, doubling the dimensionality of each input vector to $2K\overline{M}$.

\textbf{\textbullet\ Q-Network Architecture:} The Q-network is designed as a Multi-Layer Perceptron network of $U$ layers. The first $U-1$ of them alternate between fully-connected and rectified linear unit layers and the last one (output layer) is a fully-connected layer. The $u^{th}$ layer in the network has a stack of $A_u$ neurons. Two deep Q-networks are used for training stability \cite{Hasselt2016}.

\textbf{\textbullet\ Training Loss Function:} Given the objective of predicting the best interaction vector, having the highest achievable rate estimate, the model is trained using a regression loss function. at the $t^\rm{th}$ episode, the training is guided through minimizing the loss function, $\mathsf{MSE} \left( \overline{\bR}(t), \widehat{\bR}(t)\right)$, which is the mean-squared-error between the desired and the predicted output, $\overline{\bR}(t)$ and $\widehat{\bR}(t)$.

\section{Simulation Results}  \label{sec:Results}


In this section, we evaluate the performance of the proposed deep reinforcement learning solution. 

\begin{figure}[t]
	\centerline{\includegraphics[width=.9\columnwidth]{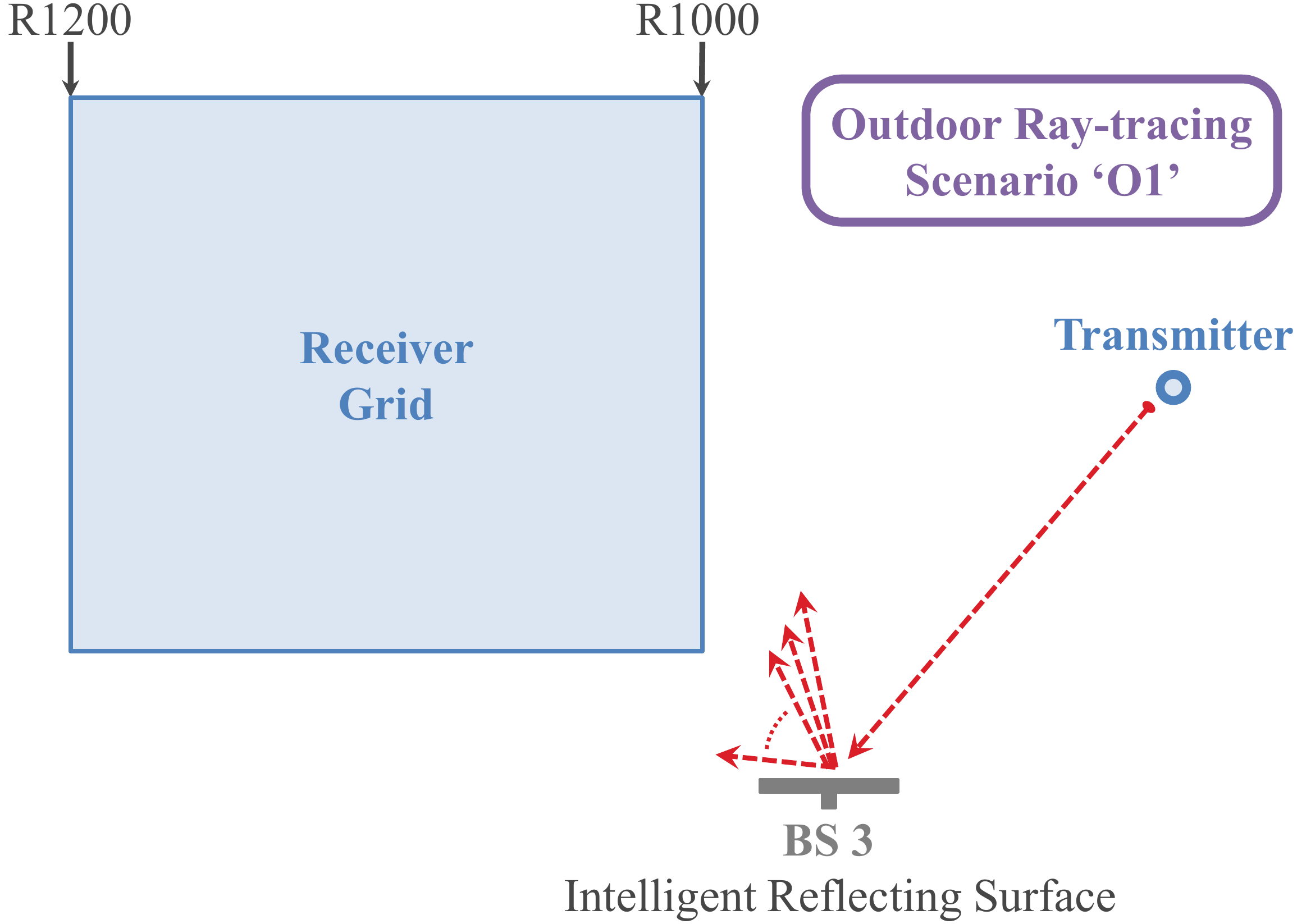}}
	\caption{The adopted ray-tracing scenario where an IRS is reflecting the signal received from one fixed transmitter to a receiver. The receiver is selected from a grid of candidate locations. This scenario is generated using Remcom Wireless InSite \cite{Remcom}, and is available on the DeepMIMO dataset \cite{DeepMIMO2019}. }
	\label{fig:Setup}
\end{figure}

\subsection{Simulation Setup} \label{subsec:Sim_Setup}
The DeepMIMO dataset in \cite{DeepMIMO2019} is adopted to generate the channels based on the outdoor ray-tracing scenario `O1'. 
The dataset parameters are summarized in Table \ref{table:DeepMIMOpset}. The transmitter's position is fixed while the receiver can take any random position in a specified x-y grid, as illustrated in \figref{fig:Setup}. We select BS $3$ to be the IRS. For a detailed description of the simulation setup, please refer to the simulation setup in \cite{Taha2019}.

\begin{table}[h]
	\caption{The adopted DeepMIMO dataset parameters}
	\begin{center}
		\begin{tabular}{ | c | c | }
			\hline
			\textbf{DeepMIMO Dataset Parameter} & \textbf{Value} \\ \hline \hline
			Frequency band &    $3.5$ GHz  \\ \hline
			Active BSs &    $3$  \\ \hline
			Active users (receivers)  &  From row R$1000$ to row R$1200$  \\ \hline
			Active user (transmitter)  &  row R$850$ column $90$  \\ \hline
			Number of BS Antennas & \begin{varwidth}[t]{\linewidth}
				{$\left(M_x, M_y, M_z\right) = \left(1,40,10 \right)$}
			\end{varwidth}
			\\ \hline
			Antenna spacing & $0.5\lambda$ \\ \hline 
			System bandwidth & $100$ MHz \\ \hline 
			Number of OFDM subcarriers & $512$ \\ \hline
			OFDM sampling factor & $1$ \\ \hline
			OFDM limit & $64$ \\ \hline 
			Number of paths & $1, 15$ \\ \hline
		\end{tabular}
	\end{center}
	\label{table:DeepMIMOpset}
\end{table}

\textbf{Deep reinforcement learning parameters:} We adopt the DRL model described in \sref{subsec:DL Model}. 
States are represented by the normalized concatenated \textit{sampled} channel of each user pair, and actions are represented by each candidate interaction vector, $\boldsymbol{\psi} \in \boldsymbol{\cP}$.
To reduce the Q-network complexity, we input the normalized \textit{sampled} channels only at the first $64$ subcarriers. The neural network architecture consists of four fully-connected layers of $4096, 16384, 16384, 4096$ nodes, respectively. Given the size of the receiver x-y grid, the DRL dataset has $36200$ data points. We split this dataset into two sets: training and testing sets, with $70\%$ and $30\%$ of the points, respectively. We consider a replay buffer of $8192$ samples and a batch size of $512$ samples. $\epsilon$ starts from $0.99$ and decrease gradually by a factor of $0.5\%$ every $40$ training iterations till it reaches $0.1$. $\gamma = 0$. $R^\rm{TH}=8.9$ bps/Hz is set to the min-max rate of the dataset.

\subsection{Achievable Rates with Deep Reinforcement Learning} \label{subsec:Compare_35_28}

\begin{figure}[t] 
	\centerline{\includegraphics[width=0.95\columnwidth,height=195pt]{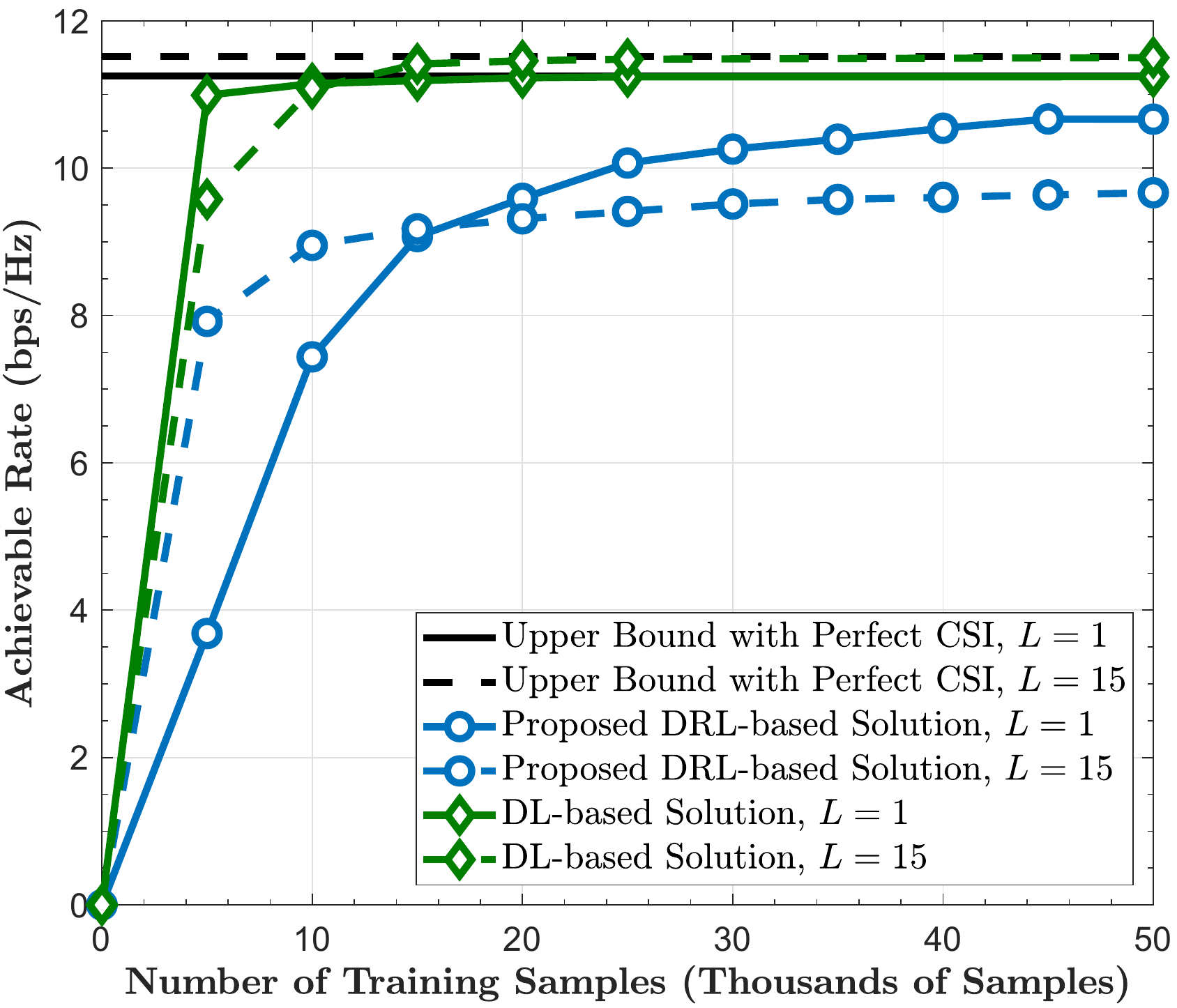}}
	\caption{The achievable rates of both the proposed deep reinforcement learning (DRL) solution and the supervised deep learning (DL) solution in \cite{Taha2019}, are compared to the upper bound, using $\overline{M}=4$ active elements for a $3.5$GHz scenario with $L\in\left\lbrace1,15\right\rbrace$ channel path(/s). The upper bound, $R^\star$ in \eqref{eq:optR}, assumes perfect channel knowledge. The figure shows the potential of the proposed DRL solution in approaching the optimal rate with almost no training overhead and a small fraction of the IRS elements to be active.} 
	\label{fig:DRL_DL}
\end{figure}

\figref{fig:DRL_DL} illustrates the achievable rate of both the proposed DRL based solution and the supervised deep learning (DL) based solution in \cite{Taha2019}, using 4 active elements with $L \in \left\lbrace1,15\right\rbrace$ channel paths. Their performances are compared to the upper bound with perfect full channel knowledge, calculated according to \eqref{eq:optR}. 
%
%
As shown, the proposed DRL solution is capable of approaching the optimal rate with more training samples that the one needed by the DL solution. 
%
%
\textbf{In contrast, the proposed DRL solution uses only one beam for each training episode, which constitute almost $\mathbf{0.3\%}$ of the beams used by the DL solution in the training phase ($\mathbf{400}$ beams).} 
This emphasizes the efficiency of the DRL solution in operating with almost no training overhead.

\begin{figure}[t]
	\centerline{\includegraphics[width=0.95\columnwidth,height=195pt]{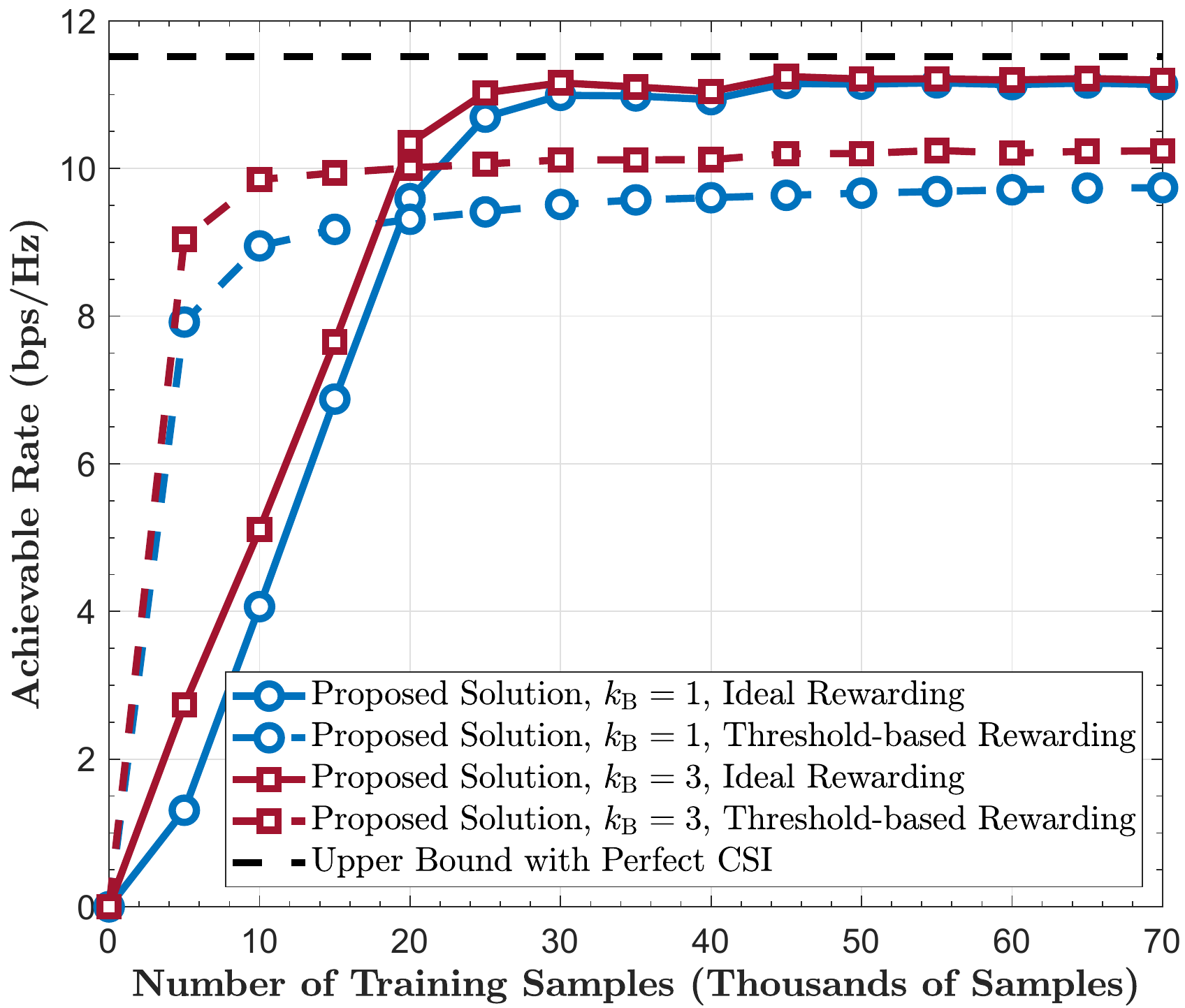}}	
	\caption{The achievable rate of the proposed DRL based approach is compared to the upper bound $R^\star$, using $\overline{M}=4$ active elements with $L=15$ channel paths. The figure illustrates the achievable rate gain when the beams selected by the deep reinforcement learning model are further refined through beam training over $k_\rm{B}$ beams.}
	\label{fig:tunning}
\end{figure}

Another candidate approach for refining the DRL prediction is to use the trained DRL model in predicting the most promising $k_\rm{B}$ beams. Then, these beams are used for beam training to identify the best beam that will be utilized for the rest of the coherence block. \figref{fig:tunning} illustrates the achievable rate of the proposed DRL based solution compared to the upper bound, at different values of $k_\rm{B}$ $(1,3)$, using 4 active elements. As demonstrated, the beam training of the promising $k_\rm{B}$ beams achieves better performance than just relying on the best network-predicted beam to reflect the incident signals. 
To test the effectiveness of the proposed framework, we examined another variant of the algorithm by updating its reward policy such that $R_\rm{Q}(t) = 1$ if $R(t) = R^\star(t)$; otherwise, $R_\rm{Q}(t) = -1$, as illustrated in \figref{fig:tunning}. The proposed DRL solution under this ideal rewarding assumption can converge to the optimal rate. This indicates that the small gap between the performance of the proposed solution and the upper bound can be explained by the practical assumptions of using threshold-based rewarding and operating in an environment with 15 channel paths.
These results shows the gains from exploring deep reinforcement learning frameworks to develop \textit{standalone} IRS architectures.

\section{Conclusion}  \label{sec:Conclusion}
For an IRS-assisted wireless communication systems, we developed an efficient solution for designing the IRS interaction matrices. 
Given an objective of designing standalone IRS architectures, the proposed solution exploits deep reinforcement learning frameworks for the IRS to learn how to predict, on its own, the optimal interaction matrices directly from the sampled channel knowledge. This solution does not require an initial dataset collection phase as opposed to the supervised learning based solutions. Simulation results based on accurate ray-tracing channels showed that the proposed solution can converge near the optimal data rates with almost no training overhead and with few active elements.
%

\balance

\end{document}

%% file: input.tex
\usepackage{amsfonts}
\usepackage{times}
\usepackage{graphicx}
\usepackage{latexsym}
\usepackage{dsfont}
\usepackage{amssymb}
\usepackage{amsmath}
\usepackage{cite}
\usepackage{verbatim}

\newtheorem{algorithm}{Algorithm}

\newcommand{\figref}[1]{{Fig.}~\ref{#1}}

\def\bb0{{\mathbb{0}}}

\def\ba{{\mathbf{a}}}
\def\bb{{\mathbf{b}}}

\def\bh{{\mathbf{h}}}

\def\bm{{\mathbf{m}}}

\def\bv{{\mathbf{v}}}
\def\bw{{\mathbf{w}}}

\def\b0{{\mathbf{0}}}

\def\bA{{\mathbf{A}}}
\def\bB{{\mathbf{B}}}

\def\bG{{\mathbf{G}}}

\def\bI{{\mathbf{I}}}

\def\bR{{\mathbf{R}}}

\def\bbC{{\mathbb{C}}}

\def\bbE{{\mathbb{E}}}

\def\bbR{{\mathbb{R}}}

\def\cD{\mathcal{D}}

\def\cN{\mathcal{N}}

\def\cP{\mathcal{P}}

\def\sf0{{\mathsf{0}}}

